\documentclass[12pt]{article}

\usepackage{dcolumn}
\usepackage{bm}
\usepackage[latin1]{inputenc}
\usepackage[spanish,english]{babel}
\usepackage{amsfonts}
\usepackage{amssymb}
\usepackage{graphicx}
\usepackage{setspace}

\newcommand{\be}{\begin{equation}}
\newcommand{\ee}{\end{equation}}
\newcommand{\bea}{\begin{eqnarray}}
\newcommand{\eea}{\end{eqnarray}}

\begin{document}

\begin{titlepage}

\begin{center}
{\large \bf Nonsingular charged black holes \`{a} la Palatini } \\
\end{center}

\begin{center}
Gonzalo J. Olmo\footnote{ gonzalo.olmo@csic.es} \\

\footnotesize \noindent {\it Departamento de F\'{i}sica Te\'{o}rica and
IFIC,  Universidad de Valencia-CSIC, Facultad de F\'{i}sica,
Burjassot-46100, Valencia, Spain.}
\end{center}

\begin{center}
Diego Rubiera-Garc\'{i}a\footnote{rubieradiego@gmail.com}\\

\footnotesize \noindent {\it Departamento de F\'{i}sica, Universidad de Oviedo, Avenida Calvo Sotelo 18, 33007, Oviedo, Asturias, Spain}
\end{center}

\begin{abstract}
We argue that the quantum nature of matter and gravity should lead to a discretization of the allowed states of the matter confined in the interior of black holes. To support and illustrate this idea, we consider a quadratic extension of General Relativity formulated \`{a} la Palatini and show that nonrotating, electrically charged black holes develop a compact core at the Planck density which is nonsingular if the mass spectrum satisfies a certain discreteness condition. We also find that the area of the core is proportional to the number of charges times the Planck area.
\end{abstract}

\small


\end{titlepage}

\section{Introduction}	

General relativity (GR) is an elegant theory that combines physics with geometry. Its field equations encode the way matter determines the causal structure of space-time, and predict the existence of black holes: regions of space-time dressed with an event horizon from which nothing can escape. Despite its experimental success, there are deep reasons to believe that GR cannot be accurate in the innermost regions of a black hole, where a space-time singularity dwells. In fact, since the laws that govern the matter fields are written on top of a curved background, when the background breaks down at a singularity, space-time can no longer tell matter how to move, matter fields can no longer talk to one another, and the laws of physics lose their predictive power.

How could singularities be avoided and how would that affect black hole structure? A necessary (but not sufficient) condition is by allowing matter to pile up forming an ultracompact core of finite density rather than concentrating on a zero volume point. This structure, however, is impossible in a typical macroscopic Schwarzschild black hole because its causal structure forces everything within the horizon to fall down towards the $r=0$ singularity. Assuming that quantum gravitational effects do not spoil the geometric nature of gravitation, in order to allow matter to pile up in the interior of the black hole one would expect that $r=$constant worldlines should recover the timelike behavior they had in the exterior region. Technically, this requires a second (inner) horizon able to restore the timelike character of the Killing vector $\partial_t$ in the innermost regions of the black hole. Quantum gravitational effects should then provide the necessary pressure or mechanism to counterbalance the gravitational pull and yield a nonsingular core of finite volume. In analogy with the physics of white dwarfs and neutron stars, one would expect this core to have a density of Planck order $\rho_P\equiv c^5/\hbar G^2\sim 10^{94}$ g/cm$^3$ \cite{Shapiro-Teukolsky}. Given this exceedingly large value, one could expect that the forces involved in the avoidance of the singularity rather than leading to a stable core could end up producing a violent and explosive phenomenon. This possibility, however, must be highly suppressed already at the classical level by the peculiar causal structure of the space-time region contained within the two horizons. There, all forms of matter and radiation are pushed towards the inner horizon, forcing the confinement of matter within the innermost region. As a result, the quantum nature of matter together with the confinement forced by gravity should constrain the allowed core configurations in much the same way a potential well constrains the states of a massive particle in nonrelativistic quantum mechanics. This could lead to a discretization of the nonsingular solutions, which should have an impact on the structure of the inner and outer horizons and/or on the mass contained within the core.

The structure with two horizons suggested above already arises in the Reissner-Nordstr\"om black holes of GR when charge and mass satisfy a certain inequality. Though in those solutions the electric repulsion is unable by itself to cure the $r=0$ singularity, some of the basic geometrical elements required to form a stable core are already present. Consequently, electrically charged black holes are a good starting point to explore the effects of new physics at the Planck scale on black hole structure. Since this new physics should provide a quantum gravitational mechanism to stabilize the matter inside the black hole, extensions of the Einstein-Hilbert Lagrangian such as
\begin{equation} \label{eq:theory}
R\to R+l_P^2(aR^2+R_{\mu\nu}R^{\mu\nu}) \ ,
\end{equation}
where  $l_P\equiv \sqrt{\hbar G/c^3} \sim 10^{-35} m$ is the Planck length, are natural candidates from an effective field theory perspective \cite{parker-toms_B&D1,parker-toms_B&D2,parker-toms_B&D3}. These extensions, however, usually introduce multiplicity of new solutions that depend on {\it ad hoc} boundary conditions, and may develop dynamical instabilities due to the higher-order character of the resulting field equations. To avoid this, one can formulate the theory  \`{a} la Palatini, i.e., assuming that metric and connection are independent entities. Though this reinterpretation of the theory is innocuous for GR, it has deep practical and conceptual implications for extensions of it because metric and affine structures are allowed to evolve independently. As a result the Palatini theory (\ref{eq:theory}) yields second-order equations \cite{Olmo2011a} and, therefore, its solutions can be completely characterized using the same parameters as in GR. Moreover, due to the peculiar
way the matter enters in the construction of the connection, the field equations exactly boil down to those of GR in vacuum but are different whenever the energy-momentum density is not zero. Therefore, the Palatini approach allows to explore the effects of Planckian physics on black hole structure in a novel way that minimizes the undesired {\it collateral effects} usually brought up by extensions of GR with higher-order derivatives and/or extra dimensions.

\section{Space-time metric and curvature}

The extension of the Reissner-Nordstr\"om solution for the Palatini Lagrangian (\ref{eq:theory}) has been found recently in exact analytical form \cite{Olmo-Rubiera}. Using Eddington-Finkelstein coordinates, the line element can be written as $ds^2=-B(r)dv^2+ 2dv dr^*+r^2(r^*)d\Omega^2$, where  $v=t+x$, $(dr/dx)^2=\sigma_- B^2$,  $(dr/dr^*)^2=\sigma_- $, $\sigma_{\pm}=1\pm1/z^4$, $z=r/r_c$ ,  $r_c=\sqrt{r_q l_P}$,  $r_q=q \sqrt{2G} $, $q$ is the electric charge, and
\begin{equation}
B(r)=\frac{1}{\sigma_+}\left(1-\frac{\left[1+\delta_1 G(z)\right]}{\delta_2 z \sigma_-^{1/2}}\right) \ .
\end{equation}
In the definition of $B(r)$ we used the dimensionless ratios ($r_S=2M=$constant)
\begin{equation}\label{eq:d1d2}
\delta_1=\frac{1}{2r_S}\sqrt{\frac{r_q^3}{l_P}} \ , \
\delta_2= \frac{\sqrt{r_q l_P} }{r_S} \ .
\end{equation}
The function $G(z)$ has a purely electrostatic origin and satisfies $\frac{dG}{dz}=\frac{z^4+1}{z^4\sqrt{z^4-1}}$, which recovers the GR value $\frac{dG^{GR}}{dz}=1/z^2$ when $z\gg 1$. Deviations from GR arise as we approach the region $z\to 1$, where $dG/dz$ diverges. This divergence manifests that the electric field cannot penetrate into the region $z<1$ (or $r<r_c$) \cite{Olmo-Rubiera}.

Expanding $G(z)$ in power series, the geometry in the far region, $z\gg 1$, is essentially that of GR, with the external horizon located at $r_+\approx \left({r_S}+\sqrt{{r^2_S}-2r_q^2}\right)/2 $ and with ${R^\alpha}_{\beta\mu\nu}  {R_\alpha}^{\beta\mu\nu}\approx \frac{12 r_S^2}{r^6}-\frac{24 r_S r_q^2}{r^7}+\frac{14 r_q^4}{r^8}+\frac{144 r_S r_c^4}{r^9}+\ldots$ However, as we approach $z\to 1$ the effects of the new physics come up and turn the function $B(r)$ into
\begin{equation}
B(r) \approx  - \frac{\left(1-\delta _1/\delta_1^*\right)}{4\delta _2}\left(\frac{1}{ \sqrt{z-1}}+O(\sqrt{z-1})\right)+\frac{1}{2}\left(1-\frac{\delta _1}{ \delta _2}\right)+O (z-1)+\ldots \label{eq:gtt_series}  \ ,
\end{equation}
where $\delta_1^*=-1/G(z)|_{z=1}\approx  0.572$. With this result, one finds  ${R^\alpha}_{\beta \mu\nu}{R_\alpha}^{\beta \mu\nu}=K_0+(1-\delta_1/\delta_1^*)K_1+(1-\delta_1/\delta_1^*)^2K_2$, with $K_1\sim 1/(z-1)^{3/2}$, $K_2\sim 1/(z-1)^3$, and $K_0$ finite at $z=1$. This implies that there is a singularity at $z=1$ except when the ratio $\delta_1$ is set to the special value $\delta_1^*$. In the other cases, $\delta_1\neq \delta_1^*$, the singularity at $z=1$ is much weaker than that found for charged black holes in GR, where the divergence at $r=0$ scales as $\sim 1/r^8$. We also note that the structure of horizons when $\delta_1<\delta_1^*$ is similar to that of a Schwarzschild black hole, whereas $\delta_1>\delta_1^*$ presents a structure closely resembling that of the usual Reissner-Nordstr\"{o}m black holes of GR.

\section{Physical aspects}

To extract the physics behind these results, we express the total charge in multiples of the electron charge, $e$, as $q=N_q e$. This allows to write $r_q=\sqrt{2\alpha_{em}}N_q l_P$, where $\alpha_{em}$ is the fine structure constant, and put the area of the hypersurface $z=1$ as $A_{c}=4\pi r_c^2= N_q\sqrt{2\alpha_{em}} A_P$,  where $A_P=4\pi l_P^2$ is the Planck area. This suggests that each charge sourcing the electric field has associated an elementary quantum of area of magnitude $\sqrt{2\alpha_{em}} A_P$. Additionally, the ratio of the total charge $q$ by the area $A_c$ turns out to be a universal constant,  $\rho_q=q/A_c=(4\pi\sqrt{2})^{-1}\sqrt{c^7/(\hbar G^2)}$, which coincides with the Planck surface charge density (up to a factor $\sqrt{2}$). Focusing on the nonsingular solutions, we find that the condition $\delta_1=\delta_1^*$ sets the following mass-to-charge relation
\begin{equation}\label{eq:rq-rs}
r_S=\frac{1}{2\delta_1^*}\sqrt{\frac{r_q^3}{l_P}} \ \leftrightarrow  \  \frac{M}{(r_ql_P)^{3/2}}=\frac{1}{4\delta_1^*}\frac{m_P}{l_P^3}\ ,
\end{equation}
where $m_P$ is the Planck mass. This indicates that the matter density inside the sphere of radius $r_{c}=\sqrt{r_ql_P}$ is another universal constant,  $\rho_{c}^*={M}/{V_{c}}|_{\delta_1=\delta_1^*}={\rho_P}/{4\delta_1^*}$. These results support the view that $z=1$ represents the surface of the black hole core, an object containing the charge $q$ uniformly distributed on its surface and with the mass $M$ in its interior at a density $\rho_{c}^*$.
This is further supported by the fact that $r=r_c$ ($z=1$) is a null surface, i.e., an inner horizon which, as reasoned above, is a necessary condition to allow for the existence of stable cores. Though the geometry in the region $z<1$ is model dependent, because it requires the specification of the matter sources carrying the mass of the core and the charge generating the external electric field, the fact that the hypersurface $z=1$ is null suggests that the interior distribution could be described by some kind of fluid whose equation of state as $z\to 1$ approaches that of radiation. An explicit realization of such a fluid could consist on a series of polytropes characterizing the matter at different densities that smoothly connect with an atmosphere of radiation as $z\to 1$. This fluid would allow to define a new auxiliary metric $\tilde{h}_{\mu\nu}$ in $z\le 1$ able to parameterize the interior of the core.  Since this is a complicated problem that requires numerical investigation,
we do not further explore it here. We also note that the fact that $A_c$ grows linearly with $N_q$ and that $\rho_q$ and $\rho_{c}^*$ only depend on $\hbar$, $G$ and $c$, and are independent of $q$ and $M$ puts forward the special quantum-gravitational nature of the nonsingular solutions.

From the regularity condition (\ref{eq:rq-rs}), we find that for a solar mass black hole, with $N_{p,\odot}\sim 10^{57}$ protons, it takes just $N_{q,\odot}=(2r_S \delta_1^* /l_P)^{2/3}/\sqrt{2\alpha_{em}}\approx 2.91 \times 10^{26}$  charges to avoid the core singularity. The ratio $N_{q,\odot}/N_{p,\odot}\sim 10^{-31}$ is so small that it is tempting to believe that such a low cost mechanism could indeed be chosen by Nature to cure black hole singularities.

On the other hand, from (\ref{eq:rq-rs}) we find that the mass spectrum of these nonsingular objects is given by $M\approx\frac{N_q^{3/2}m_P}{55}$, which takes discrete values if $N_q$ is regarded as an integer number. The discretization of the black hole mass spectrum has been found to be a very fundamental requirement for the consistency of any field theory with asymptotic Poincare-invariance because otherwise matter loops involving the contribution of virtual black holes would lead to infinities in the resulting black hole production rates \cite{Dvali10a,Dvali10b}.  Additionally, it can be shown that for  $N_q<\sqrt{2/\alpha_{em}}\sim 16.55$ there is no external horizon and, therefore, such objects can be regarded as {\it naked} cores, which have no analogue in GR. As the cores have vanishing surface gravity, these naked cores appear as natural candidates for the final state of black hole evaporation via Hawking radiation \cite{Hawking1,Hawking2,Hawking3,Fabbri:2005mw}. In this sense, the implications that the discretization of $M$ and $A_c$ could have for the Hawking evaporation of the nonsingular solutions would require a precise understanding of the connection between the quantum properties of the external (perturbative) and internal (non-perturbative) horizons.

\section{Conclusions}

Summarizing, we have argued that the quantum nature of matter and gravity should lead to a discretization of the allowed states of the matter confined in the interior of black holes.  Using a quadratic extension of GR formulated \`{a} la Palatini, we have shown that nonsingular black hole solutions with a discrete spectrum exist. Following Bohr's lessons on the Hydrogen atom, we believe that this discretized set of nonsingular solutions should be postulated as representing the physical branch of the theory, thus providing a glimpse of the effects that quantum gravity could have on the structure of black holes.

\section*{Acknowledgments}
Work supported by the Spanish grants FIS2008-06078-C03-02, FIS2011-29813-C02-02, the Consolider Program CPAN (CSD2007-00042), and the JAE-doc program of the Spanish Research Council (CSIC).

\end{document}